\documentclass[twocolumn,nobibnotes,showpacs,preprintnumbers,amsmath,amssymb]{revtex4}
\usepackage{graphicx}% Include figure files
\usepackage{dcolumn}% Align table columns on decimal point
\usepackage{bm}% bold math
\newcommand{\keV}{~\mathrm{keV}}
\newcommand{\MeV}{~\mathrm{MeV}}
\newcommand{\GeV}{~\mathrm{GeV}}
\newcommand{\TeV}{~\mathrm{TeV}}

\renewcommand{\r}{\rangle}
\renewcommand{\l}{\langle}

\newcommand{\uoned}{U(1)_d}

\newcommand{\be}{\begin{eqnarray}}
\newcommand{\ee}{\end{eqnarray}}

\newcommand {\unit} [1] {\textrm{ #1}}
\newcommand{\ald}{\alpha_d}

\begin{document}

%\preprint{APS/123-QED}

\title{Kinetic Mixing as the Origin of Light Dark Scales}% Force line breaks with \\

\author{Clifford Cheung$^{(a,b)}$, Joshua T. Ruderman$^{(c)}$, Lian-Tao Wang$^{(c)}$ and Itay Yavin$^{(c)}$}

\affiliation{(a) School of Natural Sciences, Institute for Advanced Study, Princeton, NJ 08540 \\ (b) Department of Physics, Harvard University, Cambridge, MA 02138 \\ (c) Department of Physics, Princeton University, Princeton, NJ 08544}%

\date{\today}

\begin{abstract}
We propose a model in which supersymmetric weak scale dark matter
is charged under a $U(1)_d$ dark gauge symmetry. Kinetic mixing
between $U(1)_d$ and hypercharge generates the appropriate
hierarchy of scales needed to explain PAMELA and ATIC with a GeV
scale force carrier and DAMA (or INTEGRAL) using the proposals of
inelastic (or, respectively, exciting) dark matter. Because of the extreme
simplicity of this setup, observational constraints lead to
unambiguous determination of the model parameters. In particular, the DAMA
scattering cross section is directly related to the  size of the hypercharge
D-term vacuum expectation value. The known relic abundance of DM can be used to fix the ratio of the dark sector coupling to the dark matter mass. Finally, the recent observation of cosmic ray positron and
electron excesses can be used to fix the mass of the dark matter through the observation of a shoulder in the spectrum and the size of the kinetic
mixing by fitting to the rate. These parameters can be used to make further predictions,
which can be checked at future direct detection, indirect detection,
as well as collider experiments.

\end{abstract}

\pacs{12.60.Jv, 12.60.Cn, 12.60.Fr}
\maketitle

\section{\label{sec:intro}Introduction}

A number of intriguing results from astronomical and cosmic ray data may be evidence for dark matter (DM) annihilation in
our galaxy \cite{pamela, atic}. In addition, the DAMA direct detection experiment reports a signal which may be the first instance of DM  interaction with normal matter~\cite{dama}. Interestingly, if interpreted as coming from dark matter interactions and annihilations these signals span an enormous hierarchy of length scales, 100 keV - 1
TeV, making dark matter model building a challenging enterprise.
Along these lines, Arkani-Hamed et$.$ al$.$ have suggested a broad
framework \cite{nima_paper} in which 500 - 800 GeV dark matter is
charged under a dark gauge group, $G_d\supset U(1)_d$, whose
abelian factor kinetically mixes with the Standard Model (SM)
photon \cite{Pospelov:2007mp}. Assuming that $G_d$ is non-abelian and broken at
a GeV, then loops of the resulting GeV scale dark gauge bosons
will generate 100 keV - 1 MeV mass splittings within the dark
matter multiplet.

This approach resolves several of the puzzles raised by recent
observations: 1) a Sommerfeld enhancement from GeV scale dark
gauge bosons boosts the dark matter annihilation rate today,
reconciling the large flux of $e^+e^-$ observed in the PAMELA and ATIC data
with the weak cross-section inferred from the dark matter relic
abundance. 2) Dark gauge bosons have kinematically suppressed
hadronic decays, explaining the lack of excess anti-protons at
PAMELA\@. 3) The DAMA signal arises from inelastic nuclei-dark
matter (iDM) scattering coming from $\sim 100\keV$ mass splittings
in the dark matter multiplet \cite{idm}. 4) A slightly larger
splitting of $\sim 1 \MeV$ can explain the INTEGRAL 511 keV line
using the proposal of exciting dark matter (XDM) \cite{xdm, cgs}.

In this paper, we propose a simple and predictive setup that
differs from the original proposal in two important ways. First,
we demonstrate that an abelian gauge group, $G_d = U(1)_d$, is
sufficient to generate a multiplet of states with 100 keV - 1 MeV
mass splittings.  This a considerable simplification over
non-abelian models, which generally need large numbers of dark
Higgses \cite{na_ds}.  Second, we argue that supersymmetric
theories acquire a scale of $\sim$ GeV as a dynamical consequence
of kinetic mixing alone. D-term mixing between $U(1)_d$ and the
Minimal Supersymmetric SM (MSSM) hypercharge induces an effective
Fayet-Iliopoulos (FI) term for $\uoned$ of order GeV$^2$. In the
presence of a single $U(1)_d$ charged dark Higgs, the dark gauge
symmetry is broken at a GeV\@.

Our model is also distinct from a number of alternative proposals
for $U(1)_d$ charged dark matter.   An earlier treatment has
neglected D-term mixing completely, even though it is dominant to
the effects being considered \cite{Zurek}.  Another approach
considers gauge mediation in the limit of negligible kinetic
mixing \cite{ChunPark}. Also, during the completion of this work
we learned of an interesting forthcoming proposal \cite{Raman}
which utilizes anomaly mediation to generate additional GeV scale
contributions.

The model presented in this paper is remarkably simple and has a
small number of free parameters that can be fixed by observations.
For instance, the dark matter mass, $M$, may be fixed by a possible future observation of a shoulder in the PAMELA positron excess (alternatively, the
feature seen in ATIC may already fix that scale). Assuming thermal
freeze-out, the dark gauge coupling, $g_{d}$, determines the relic
abundance and is thus fixed in terms of $M$. Furthermore, the size
of the kinetic mixing, $\epsilon$, is constrained by the boost
factor required to explain the PAMELA positron excess. Quite
interestingly, our model implies a direct relation between the
DAMA scattering cross-section and the hypercharge D-term of the
MSSM\@.  This is a generic prediction of any theory whose only
origin of scales is the kinetic mixing. Also, the dark photon
should be copiously produced in SUSY cascades at particle
colliders, yielding clusters of closely packed collimated
energetic leptons, which has been referred to in Ref.~\cite{nima_paper,na_ds} as ``lepton
jets''. At high energy colliders, dark photons are
produced with energies of 10s $-$ 100s GeV.  As the dark photon decays
through its mixing with the photon it produces a pair of leptons that are
highly collimated $\Delta R_{\ell \ell} \sim 0.1 - 0.01$,  forming 
lepton jets. Cascade decays in the GeV dark sector will result in more leptons and hence richer lepton jets, see \cite{na_ds} for a detailed discussion. We note that the dark photon can also have  significant decay branching ratios to light mesons. Presence of such mesons certainly
alters the phenomenology, and it is therefore possible to define several
sub-classes of lepton jets. Since our focus is not on
the collider phenomenology, we will not make this distinction in this
paper.

\section{New Scales in the Dark Sector via Kinetic Mixing}

Throughout this letter we assume a weak scale mass
for the dark matter supermultiplet. All of the light scales in the
dark sector will be generated dynamically through kinetic mixing
between hypercharge and the dark force carrier.  In particular,
the action contains a term
\begin{equation}
\label{eqn:WWmix}
\mathcal{L} \supset -\frac{\epsilon}{2}\int d^2\theta W_Y W_d
\end{equation}
where $W_Y$ ($W_d$) is the supersymmetric field strength for the
SM hypercharge (dark abelian group) and $\epsilon$ is a small parameter. Integrating out heavy fields charged under both hypercharge and $\uoned$ will induce this operator and we can estimate the size of the mixing to be,
\be \epsilon &=& -\frac{g_Y g_y}{16\pi^2} \sum_i Q_{i} q_{i}
\log\left(\frac{M_i^2}{\mu^2}\right) \ee
which can naturally be of order $10^{-3}-10^{-4}$. As observed in ref. \cite{na_ds}, the kinetic mixing includes
\be V_{\textrm{D-term mixing}} &=& \epsilon D_Y D_d \ee
After electroweak symmetry breaking, $D_Y$ gets a vacuum expectation value (VEV)
\be
\label{eqn:DYVEV}
\l D_Y \r &=& \frac{g'}{2}\left(|H_u|^2 - |H_d|^2\right) +
\xi_Y \ee
where $H_{u,d}$ are the MSSM Higgs doublets and $g'$ is the hypercharge coupling. We included, $\xi_Y$, which is an effective FI term for the SM hypercharge group since this is a relevant
operator allowed by all the symmetries and there is no reason to a
priori exclude it from the low energy action that defines the MSSM
\cite{deGouvea:1998yp, Dimopoulos:1996ig, anomalyFI}. From the low energy perspective, $\xi_Y$ is only constrained to not be so large as to destabilize the electroweak scale.

The expectation value of $D_Y$ induces an effective FI term for
the dark abelian group via the kinetic mixing
\be \label{eqn:xi}
 \xi &=& \epsilon \l D_Y \r = \epsilon\left(- \frac{g'  v^2 \cos 2\beta}{4} + \xi_Y\right)
 %\\\nonumber &=& - 0.54
%\textrm{ GeV}^2 \cos 2\beta \left(\frac{\epsilon}{10^{-4}}\right)
\ee
where $v$ is the electroweak vacuum expectation value. With
$\epsilon = 10^{-4} -10^{-3}$ and $\l D_Y \r$ of order the weak
scale, we find that $\xi = (1 - 5 \unit{GeV})^2$. Thus the GeV
scale in the dark sector is a fortuitous byproduct of the kinetic
mixing. If there are any light degrees of freedom charged under
$\uoned$, the vacuum can break the gauge group and/or
supersymmetry at the GeV scale.  We focus on the possibility that
SUSY is preserved and $\uoned$ is broken, resulting in a
$\sim\GeV$ mass for the vector supermultiplet.  If dark matter has
a superpotential coupling to a light field that gets a VEV, then
the vacuum can dynamically generate an MeV sized splitting between
dark matter supermultiplets, as we now demonstrate with a concrete
model.

\section{A Model for the Dark Sector}

We take DM to be a pair of chiral supermultiplets, $(\Phi,\Phi^c)$, oppositely
charged under the dark gauge group, $U(1)_d$.  The superpotential for chiral multiplets is given by,
\be
\label{eqn:W} W &=& M \Phi \Phi^c + \lambda N h h^c + \frac{1}{4\Lambda}
\Phi^2 h^{c2}
\ee
$h$ and $h^c$ are oppositely charged and $N$ is neutral under
$\uoned$.  Here $\Lambda \sim \TeV$ and is associated with
new physics at the electroweak scale \footnote{This can come in the form of TeV scale
  states which couple to both $\Phi$ and $h^c$. Then the operator $\frac{1}{4\Lambda}
\Phi^2 h^{c2}$ can be generated after integrating out the heavy states. One
simple possibility is to introduce an additional pair of
chiral supermultiplets ${S, S^{c}}$ with Dirac mass $\sim$ TeV, and 
couplings $\lambda_{\Phi} S \Phi^2 + \lambda_{h^c} S^c h^{c2}$.}.  We choose a discrete
symmetry to forbid dimensionful operators involving the light
fields: $N$, $N^2$, and $h h^c$.  Depending on the choice of
discrete symmetry, there may be marginal and irrelevant operators
in addition to the ones included in equation \ref{eqn:W}, but they
will not be relevant to the following discussion.  The neutral
field is included in order to avoid any massless degrees of
freedom at low energies.

The scalar potential for this theory is given by:
\begin{eqnarray} V &=& V_D + V_F \nonumber \\
V_D &=& \frac{1}{2}\left[\frac{g_d}{2}(|h|^2-|h^c|^2 +
|\Phi|^2 -|\Phi^c|^2)+\xi \right]^2 \nonumber \\
V_F &=& \left|M \Phi^c + \frac{1}{2\Lambda} \Phi h^{c2}\right|^2  +\left|M \Phi \right|^2 \nonumber \\ &+&
\left| \lambda N h + \frac{1}{2\Lambda} h^c \Phi^2 \right|^2 + \left|\lambda Nh^c\right|^2 + \left|\lambda h h^c \right|^2
\end{eqnarray}
There is a SUSY vacuum (with vanishing F and D-terms) with broken $\uoned$ at $\l h^c \r =\sqrt{\frac{2\xi}{g_d}}$ and all other scalar VEVs set to zero.

\subsection{Coupling to the Standard Model}
The supersymmetric field strength mixing, Eq.~(\ref{eqn:WWmix}), also contains the gauge-boson kinetic mixing,
\be \mathcal{L}_{\textrm{gauge mixing}} &=& \frac{\epsilon}{2}
B_{\mu\nu} b^{\mu\nu} \ee
where $B_{\mu\nu}$ and $b_{\mu\nu}$ denote the SM and dark
field strengths. As argued in \cite{nima_paper}, this
is the leading marginal operator that couples the dark sector to
the standard model.  Moreover, since it does not violate any SM
symmetries, it is relatively unconstrained phenomenologically, and $\epsilon = 10^{-4}
-10^{-3}$ is consistent with current bounds \cite{Pospelov:2008zw}.

The primary effect of this mixing is to induce an $\epsilon$
suppressed coupling between the electromagnetic current of the SM
and the dark vector-boson \cite{Holdom:1985ag}. Dark matter
annihilations produce dark vector-bosons that subsequently decay
into light leptons or pions via this kinetic mixing. Decay into heavier particles is kinematically disallowed. This injection of leptons can explain the excesses observed at PAMELA and ATIC\@.  Another
consequence of the kinetic mixing is that the SM $Z$ boson and
bino couple to the $\uoned$ current, which has implications for
collider phenomenology \cite{na_ds}.

\subsection{Mass Spectrum}

Next, let us consider the spectrum of the dark sector. The vector supermultiplet gets a mass of
\be
 \label{eqn:mb}
 m_b^2 &=& g_d \xi
 \ee
which is naturally $\sim\GeV$ scale. As pointed out in Refs. \cite{cgs,nima_paper}, a vector-boson of this mass elegantly explains why decay channels into anti-protons are kinematically disfavored. It also serves as a light force carrier capable of enhancing the annihilation cross-section via the Sommerfeld enhancement mechanism to produce the necessary annihilation rate for PAMELA or ATIC\@.

The masses of the dark matter states $\Phi$ and $\Phi^c$ are affected as well. To leading order in $m_b/M \lesssim 10^{-3}$, the mass eigenstates are given by $\Phi_\pm = (\Phi \pm
\Phi^c)/\sqrt{2}$, and $\Phi_-$, being the lighter of the two, is identified with our (fully supersymmetric) dark matter candidate. The mass splitting between the two states is,
\be
\Delta m &=& m_+ - m_- = \frac{m_b^2}{g_d^2 \Lambda} \\
&=& 0.8 \MeV \left(\frac{30^{-1}}{\ald} \right)\left(\frac{m_b}{\GeV}\right)^2 \left(\frac{3\TeV}{\Lambda}\right) \nonumber
\ee
Depending on the precise values of the parameters involved, this
scale can be used to explain either DAMA or INTEGRAL using the iDM
or XDM scenarios, respectively. Either way, the mass splitting can be fixed by fitting to the experimental signature Furthermore, since the states are
almost maximally mixed, the transitions among mass eigenstates are
strongly inelastic. The elastic couplings are suppressed by
$m_b^2/M\Lambda$ and bounds from direct detection are easily evaded. If $\Delta m < 2 m_{electron}$, then  the life-time of the excited state is longer than the age of the universe\footnote{We thank N. Arkani-Hamed and M. Pospelov for pointing this out to us. See Ref. \cite{Finkbeiner:2009mi} for details and lifetime estimates.}. Direct detection bounds may be relevant if a cosmological abundance of the excited state is still present today.

Lastly, we consider the spectrum of dark Higgses.  $h^c$ is eaten
via the super-Higgs mechanism. $h$ and $N$ pair-up and become massive:
\be \label{eqn:mHmS} m_h^2 = m_N^2 = \frac{2 \lambda^2 \xi}{g_d} = 2
\left(\frac{\lambda}{g_d}\right)^2 m_b^2 \ee
As long as $\lambda > g_d/2\sqrt{2}$, then $m_h > m_b /2$ and
decays of the dark photon into dark Higgses are kinematically
forbidden. At this level, these states are exactly stable since the potential respects $N-h$ number as an exact symmetry.

\begin{figure}[h]
\includegraphics[scale=.31]{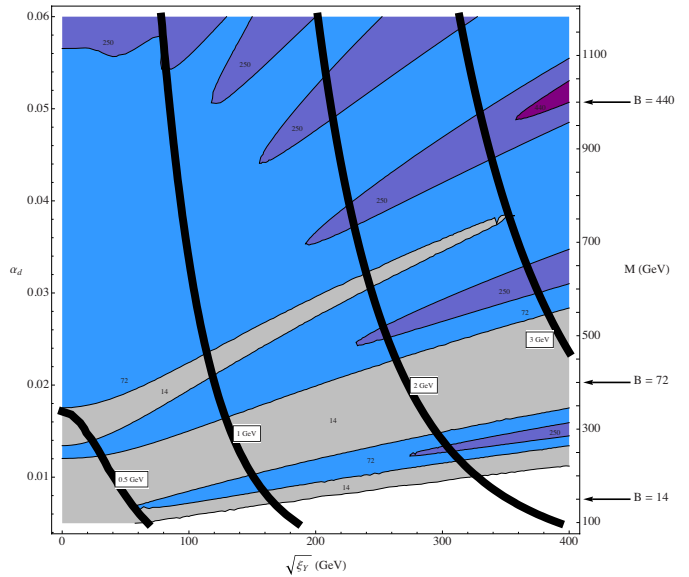}
\vspace{ -0.8 cm}
 \caption{A contour plot of the Sommerfeld enhancement
\cite{nima_paper} as a function of
$M$ (or $\ald$, related through Eq.~(\ref{eqn:freezeCS})) and $\sqrt{\xi_Y}$, with $\epsilon=10^{-4}$ and $\tan \beta = 40$.  The solid black lines correspond to fixed $m_b$.  The dark matter velocity is taken to be $v =150 \unit{km/s}$.  On the right we indicate the boost required for PAMELA for three reference values of $M$, for the process $\Phi \Phi \rightarrow
\gamma^\prime \gamma^\prime$ followed by $\gamma' \rightarrow e^+ e^-$ \cite{boost}.  The boost is relative to $\l \sigma v \r = 3 \times 10^{-26} \unit{cm}^3 \unit{s}^{-1}$, assuming local dark matter density $\rho_0 = 0.3 \unit{GeV} \unit{cm}^{-3}$.
\label{fig:S10-4} }
\end{figure}

\subsection{Supersymmetry Breaking Effects}

This setup is interesting because the origin of scales is centered
on the mixed D-term.  For this reason we have assumed there are no
large SUSY breaking effects, for example from gravity mediation.
Instead, we consider low-scale gauge mediation such that the dark
sector only receives SUSY breaking contributions from the MSSM via
kinetic mixing.  As such, the dark sector is supersymmetric only
at leading order in $\epsilon$.  For example, since the dark Higgs
couples to the MSSM bino, it receives a positive soft mass of
order $m^2 \sim \epsilon^2 g_d^2 M_{\tilde B}^2 / 16 \pi^2 \sim
(10 - 100 \textrm{ MeV})^2$ from bino loops. This has a negligibly
small effect on the dark matter supermultiplet, since it lifts the
scalars from the fermions by $m^2 /
M\sim~\mathcal{O}(1)~\mathrm{keV}$.  The dark matter is
effectively supersymmetric because transitions among the
supermultiplet occur on timescales longer than the age of the
universe.

In contrast, the scalar $h$ is heavier than its superpartner by an
amount $m^2/m_h \sim\mathcal{O}(1)~ \MeV$, while the scalar $N$ is lighter by
a tenth of that due to a \textsl{negative} mass$^2$ contribution
from the Yukawa coupling, $\lambda N h h^c$. Furthermore, an
A-term of size $\lambda\epsilon^2 g'^2 M_{\tilde B} /16\pi^2$ is
generated which mixes $N$ and $h$ once $h^c$ condenses. The
resulting spectrum consists of the scalar $N$ as the lightest state,
the fermionic $N$ and $h$ about $100\keV - \MeV$ above that, and
the scalar $h$ as the heaviest state.  While these MeV splittings
imply interesting possibilities for model building, we leave this
for future work.

\section{Observations and Predictions}

In this section we discuss the parameters of our model and their
relation to observations in astrophysics and direct detection.

The mass of the WIMP can be determined through the
electron/positron excess seen in ATIC/PAMELA\@. Since the leptons
are produced as byproducts of the annihilation into dark photons,
$\Phi\Phi\rightarrow \gamma^\prime\gamma^\prime$, followed by a
leptonic decay of $\gamma^\prime$, the excess should have an
endpoint at the WIMP's mass.

The relic abundance of dark matter is fixed by the thermal
annihilation cross-section of $\Phi$ during freeze-out. The
dominant annihilation channel is $\Phi\Phi \rightarrow
\gamma^\prime \gamma^\prime$, which fixes the dark gauge coupling
in terms of the dark matter mass \footnote{We leave a more
detailed calculation of the relic abundance to a future
publication.}:
\be \label{eqn:freezeCS} \l \sigma v \r_{\rm
freeze} &\sim& \alpha_d^2/M^2
\ee
with $\l \sigma v \r_{\rm freeze} = 3 \times 10^{-26} \unit{cm}^3 \unit{s}^{-1}$.

Next, let us consider the  effects of the GeV scale states on
cosmology.  Although $N$ and $h$ are stable, their relic abundance
can be sufficiently depleted assuming that they are heavier than
the vector multiplet. In this case they annihilate efficiently
into the vector multiplet, which in turn annihilates into SM
fields via the kinetic mixing.  Since this cross-section is small,
the states in the vector multiplet have large abundances at
freeze-out, but all decay safely before Big Bang Nucleosynthesis
(BBN).  In particular, the dark photon decays promptly to $e^+
e^-$, while the radial dark Higgs decays either into $e^+ e^-$
through a loop or into $e^+e^-e^+e^-$ through two off-shell dark
photons.  The dark photino has the longest lifetime because it has
to decay to a photon and gravitino.  This decay can occur before
BBN:
\begin{equation}
\label{eqn:gravitinorate}
\tau_{\tilde{\gamma}'\rightarrow\gamma\tilde{G}} \sim 0.3\unit{s} \left( \frac{10^{-3}}{\epsilon} \right)^2
\left(\frac{\unit{GeV}}{m_b}\right)^5 \left(
\frac{\sqrt{F}}{10\unit{TeV}} \right)^4  
\end{equation}

\begin{figure}[h]
\includegraphics[scale=0.8]{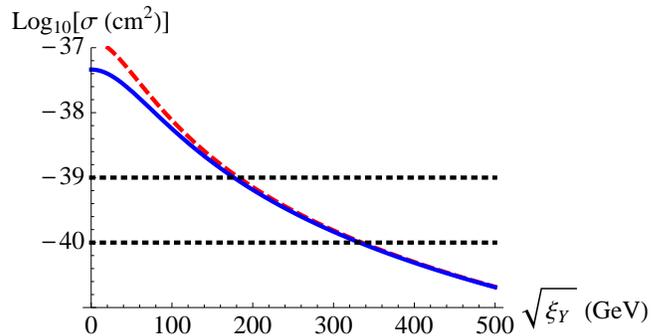}
 \caption{\label{fig:DAMAwithFI} The DAMA cross-section as a
function of the effective FI parameter with $\tan \beta = 2$ (red,
dashed) and $\infty$ (blue, line).  The horizontal lines indicate
the preferred range for the WIMP-nucleon cross-section
\cite{idm}.}
\end{figure}

The dark photon, $\gamma^\prime$, provides the Sommerfeld
enhancement of the annihilation cross-section needed for PAMELA
\cite{pamela} and ATIC \cite{atic}.  In Fig.~\ref{fig:S10-4}, we
present a contour plot of the Sommerfeld enhancement, $S$, as a
function of $\xi_Y$ and $M$.

Our model can reconcile DAMA with the limits of other direct
detection experiments, if $\Lambda$ is such that the mass
splitting between $\Phi_+$ and $\Phi_-$ is $\Delta m \sim
100\keV$, providing a realization of the iDM scenario of
Ref.~\cite{idm}. Interestingly enough, the cross-section per
nucleon in this model is sensitive to $\l D_Y \r$ only,
\be
\label{eqn:damaxs}
\sigma = \frac{4 Z^2 \mu^2_{ne}}{A^2} \frac{\alpha \cos^2
\theta_W}{\left< D_Y \right>^2}
\ee
Notice that any dependence on the dark sector couplings have cancelled completely.  This type
of cancellation will occur in {\it any theory in which the mass
scale of the dark sector is fixed entirely by D-term kinetic
mixing}. Thus, for this class of models the measured DAMA
cross-section, if confirmed, yields a definitive prediction about
electroweak physics. Fig. \ref{fig:DAMAwithFI} shows the
dependence of the DAMA cross-section on $\xi_Y$, where we have
also denoted the range of cross-sections preferred by
iDM~\cite{idm}.

A determination of $\xi$ and $\ald$ yields a prediction for $m_b$,
the mass of the vector supermultiplet, as shown in Fig.
\ref{fig:S10-4}.   Furthermore, since the MSSM bino couples
directly to the dark photon and photino, dark photons should be
produced in SUSY cascades from the MSSM at high energy colliders. Such
a signal, very distinct from the collider signatures of the conventional MSSM,  is generic in GeV dark sector models with supersymmetry
\cite{na_ds}. Dark state production at the LHC provides a
promising avenue for probing the dark sector and its interactions
\cite{na_ds,Bai:2009it}.

\section{Conclusions}

In this letter we considered a supersymmetric dark sector involving a massive and stable matter field which constitute the WIMP and is coupled to an abelian gauge field. The sector also contains light matter fields which ultimately spontaneously break the gauge symmetry at $\sim\GeV$. Similarly to recently proposed scenarios, the abelian group is weakly mixed with hypercharge through the kinetic terms. The main point of our discussion is that, in supersymmetry, the existence of kinetic mixing, together with the breaking of hypercharge in the SM, also breaks the dark abelian gauge symmetry at $\sim\GeV$ in a natural and unavoidable fashion. 

We find the extreme simplicity of this setup and the natural
generation of all the scales involved the most attractive feature of
this scenario. However, as it stands, it suffers from two main
difficulties. The first is related to the lifetime of the dark gaugino
given in Eq. (\ref{eqn:gravitinorate}). In order to decay before BBN
it requires a rather low supersymmetry breaking scale. This can easily
be relaxed with a slightly heavier gauge boson or larger mixing
parameter. Moreover, the injection of electromagnetic energy during
BBN is not as strongly constrained as hadronic energy
injection~\cite{Jedamzik:2006xz}. The second issue, and the more
serious one, is associated with the cross-section of WIMP-nucleon
scattering in the iDM model as given in Eq. (\ref{eqn:damaxs}). The
model's simplicity leads to an unambiguous relation between this
cross-section and the MSSM hypercharge D-term, $\left< D_Y
\right>$. In the simplest case, where $\left< D_Y \right>$ is given by
the higgs' VEV alone, the cross-section is too large by about an order
of magnitude. It is certainly possible for $\left< D_Y \right>$ to
enjoy from additional contributions, however, some of the model's
allure is  marred.  

Despite these shortcomings, both of which can be alleviated with the
slightest extensions, this model serves as an example for an
incredibly simple dark sector which \textit{naturally} generates all
the necessary scales recently discussed in the literature. It exhibits
a rich structure originating from the single inclusion of
supersymmetric kinetic mixing between hypercharge and a dark abelian
gauge group. Most importantly, it is predictive in its content and
results in unambigious relations between its parameters and measurable
quantities.

\section{Note added:}
Several months after the completion of the initial version of this
paper, Fermi-LAT published their first of measurement of the $e^+ + e^-$
spectrum, which shows a deviation from the conventional background
prediction \cite{Abdo:2009zk}. A boost factor on the order of $10^2$
is still necessary  to account for the excess from dark matter
annihilations \cite{Meade:2009iu}. The model presented in this paper
remains a promising candidate for this scenario. 

%The best fit to Fermi-LAT suggests dark
%matter mass to be 2 - 3 TeV, which is outside of our parameter scan
%presented in Fig. 1. With such a heavier dark matter, we expect our model to be viable with slightly
%larger $\ald$. and somewhat larger $m_b$.  

\begin{acknowledgments}
We would like to thank N. Arkani-Hamed and N. Weiner for very useful
discussions. We would especially like to thank T. Slatyer for help in generating the Sommerfeld enhancement plot. L.-T. W. and I. Y. are supported by the National
Science Foundation under grant PHY-0756966 and the Department of
Energy under grant DE-FG02-90ER40542.  J. T. R. is supported by a
National Science Foundation fellowship.
\end{acknowledgments}

%: bibliography

\end{document}